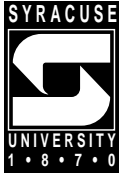

# The Renormalization Group and Quantum Hall Edge States

Varghese John$^a$, Gerard Jungman$^b$, Sachindeo Vaidya$^c$

*Department of Physics, Syracuse University, Syracuse, NY 13244*

**Abstract**

The role of edge states in phenomena like the quantum Hall effect is well known. In this paper we show how the choice of boundary conditions for a one-particle Schrödinger equation can give rise to states localized at the edge of the system. We consider both the example of a free particle and the more involved example of a particle in a magnetic field. In each case, edge states arise from a non-trivial scaling limit involving the boundary conditions. Second quantization of these quantum mechanical systems leads to a multi-particle ground state carrying a persistent current at the edge. We show that the theory quantized with this vacuum displays an "anomaly" at the edge which is the mark of a quantized Hall conductivity in the presence of an external magnetic field. We also offer interpretations for the physics of such boundary conditions which may have a bearing on the nature of the excitations of these systems.

$^a$vjohn@npac.syr.edu
$^b$jungman@npac.syr.edu
$^c$vaidya@suhep.syr.edu

## 1. Introduction

Quantum states localized at the boundaries of spatial regions are a very generic feature in theories with a gauge symmetry. Such theories describe diverse phenomena, from elementary particle physics to condensed matter systems such as the quantum Hall effect. In gauge theories, edge states occur when a system admits localized charges at the boundary, constructed from the Gauss law, appropriately smeared with test functions that do not vanish at the boundary [1][2]. Since it is believed that quantum Hall systems are described by (topological) gauge theories, edge states can play an important role in the description of quantum Hall samples on finite geometries.

In fact, there is a much simpler motivation for considering edge states in quantum Hall systems, as pointed out by Halperin [3]. Although no current is supported by the bulk states of a quantum Hall system, single-particle electronic states near the boundary do support an edge current, localized within one magnetic length of the edge. Now, if one imagines filling an appropriate Fermi sea of single-particle states, the resulting occupied states at the edge give rise to a persistent edge current. When an electric field is applied tangential to the boundary, charges flow from the bulk and accumulate at the edge. This charge transport is the Hall current [4].

In the effective Chern-Simons gauge theoretic description of the quantum Hall effect [5], the edge states are described by a conformal field theory absolutely localized at the edge [6][7]. The flow of charges from the bulk to the edge, or more specifically from one edge to another, can be described as an anomaly in the 1+1 dimensional theory at the edge [8][9]. Note that, in the context of the quantum Hall effect, the fact that the edge states in the gauge theoretic description are absolutely localized at the edge should be considered as an artifact of the long-distance limit implied in the use of Chern-Simons theory as an effective description.

The connection between the anomalous edge field theory and the simple quantum mechanical picture of edge states has, as far as we can tell, never been made explicit in the literature. In this paper we show how the choice of boundary conditions for a one-particle Schrödinger equation can give rise to states localized at the edge of the system. Because of its relative simplicity, we first discuss the case of a free particle on a disk and show how edge states can be made to appear in this case. These edge states are similar to those arising for the case of a particle in a magnetic field. Second quantization of these quantum mechanical systems leads to a multi-particle ground state which carries a persistent current at the edge. We further show that the theory quantized with this vacuum displays an "anomaly" at the edge which is the mark of a quantized Hall conductivity.

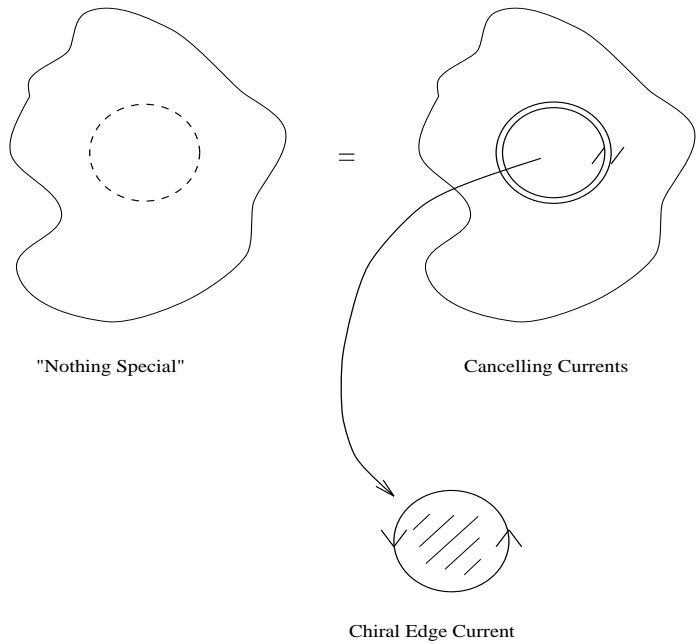

Figure 1. Cartoon of the ground state and the genesis of the edge states in a restricted region of space.

We will present several interpretations of our boundary condition, including an interpretation in terms of interaction with an edge current which provides a mechanistic derivation of the condition. At this point it is worthwhile to explain how the boundary condition came to be considered. Let $\psi$ be the Landau wave function for a particle moving in a magnetic field on an infinite plane. Suppose that we introduce a fictitious boundary in the plane without disturbing the particle, by simply drawing a circle in the plane. That this procedure should not affect the particle means that the boundary condition on this circle should be given by the behaviour of the infinite-plane wave function on that circle, with no modification. Let $\psi_{mn}$ be a (non-normalized) Landau wave function for a particle moving in a plane with a constant pependicular magnetic field $B$,

$$\psi_{mn} = e^{im\theta}\exp(-\frac{1}{4}|eB|r^2)r^{|m|/2}{}_1F_1(-n, 1+|m|, \frac{1}{2}|eB|r^2). \quad (1.1)$$

Pick some radius $R$ and consider states which have support near that radius. If that radius is much larger than the magnetic length, then these states have large angular momentum. In the limit $|m| \to \infty$, a calculation shows

$$\partial_r \psi_{mn} \sim \left[-\frac{1}{2}|eB|R + \frac{|m|}{R}\right]\psi_{nm}. \quad (1.2)$$

Now consider the energies of these states, $E_{mn} = n + |m|/2 - m\,\mathrm{sgn}(eB)/2 + 1/2$. When the angular momentum "goes along with" the magnetic field, i.e. $m\,\mathrm{sgn}(eB) > 0$, the energy is low; in fact all such states with the same $n$ are degenerate. However, states with the opposite sign of angular momentum have significantly higher energies. Therefore, the lowest energy states are of one chirality and a truncation of the theory to the lowest energy states is necessarily a chiral procedure. Therefore, our candidates for "edge-like" states are given by Eq.(1.1) with $m\,\mathrm{sgn}(eB) > 0$. Using this condition and using the relation $A_\theta = Br/2$, we can write Eq.(1.1) as $\partial_r \psi \sim -iR^{-1}D_\theta \psi$, where $D_\theta$ is the gauge-covariant derivative. A sketch of this interpretation is given in Fig. 1. The chiral edge current illustrated there arises for two reasons. First is the truncation to lower energy modes; second is the procedure of isolating some finite region.

This procedure has some formal relation to similar procedures in gauge theories. From the standpoint of an observer who makes measurements confined to the interior of such a region, there will generically arise quantum states localized at the boundary. In the Chern-Simons theory these states are generated by a chiral current algebra at the edge.

## 2. A Sticky Boundary Condition

Consider a non-relativistic particle of mass $M$ propagating on a disk of radius $R$. Take the Hamiltonian to be that of a free particle,

$$H = -\frac{1}{2M}\Delta, \quad (2.1)$$

$\Delta$ being the Laplacian on the disk. To completely specify the system, we must choose a boundary condition for the wave function of the particle. Choosing a local and linear boundary condition which has at most one derivative, we have

$$\partial_r \Psi = i\mu \frac{1}{R}\partial_\theta \Psi + \lambda(\theta)\Psi \qquad \text{on } \partial D. \tag{2.2}$$

In order to keep rotational invariance, and also for the sake of simplicity, in this Section we will assume that $\lambda(\theta) = \lambda$ is a constant; this does not change the conclusions of the analysis. In the case that $\mu = 0$, this boundary condition is a one-parameter family which interpolates between the standard Dirichlet and Neumann boundary conditions. Note that the boundary condition with $\mu \neq 0$ is inherently chiral, with the choice of chirality given by the sign of $\mu$.

The Hilbert space for this system is the space of square-integrable functions on the disk. The boundary condition defines a domain for the Hamiltonian operator, which is formally given by the Laplacian. As usual, we must check that the Hamiltonian with this domain is a self-adjoint operator, in order to have a consistent quantum-mechanical time evolution for the system. That this operator is symmetric is a simple exercise in integration by parts; note that the symmetry relies on the fact that $\mu$ is a constant. The full self-adjointness condition can be shown by following standard calculations in Ref. [10]. Physically, the criterion of self-adjointness means simply that the probability current is conserved, and the particle cannot escape from the disk.

From the rotational symmetry, the Hilbert space decomposes into sectors of fixed angular momentum, labelled by $m$. The eigenfunctions for this Hamiltonian are of the form

$$\Psi = e^{im\theta} J_m(\omega_{mn} r), \tag{2.3}$$

where $J_m$ is the Bessel function of order $m$. The eigenvalue condition following from the boundary condition of Eq.(2.2), in a given angular momentum sector, is

$$R\lambda - \mu m = R\omega_{nm}\frac{I'_m(R\omega_{nm})}{I_m(R\omega_{nm})}, \tag{2.4}$$

where $n$ labels the roots of Eq. (2.4) for fixed $m$. When $\omega_{nm}$ is real, the energy of the state is negative. We have chosen to write Eq. (2.4) in terms of the modified Bessel functions, $I_m(x) = i^m J_m(ix)$, since we will be especially interested in the negative energy states.

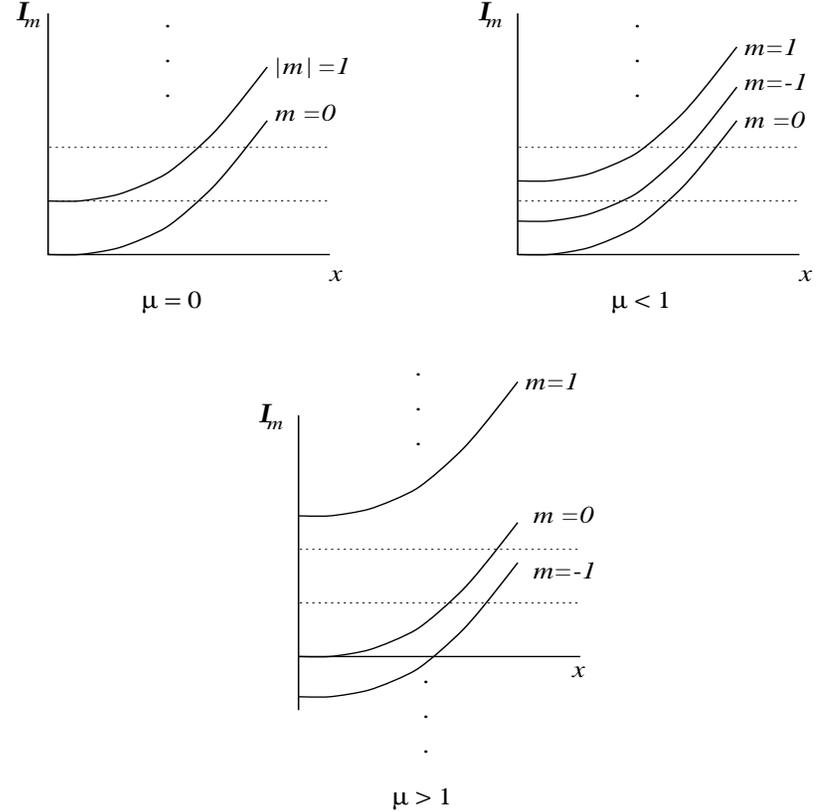

Figure 2. Graphical explanation of the eigenvalue condition which leads to energies unbounded from below.

Define the convenient function

$$\mathcal{I}_m(x) = \mu m + x I'_m(x)/I_m(x). \tag{2.5}$$

It can be shown that $\mathcal{I}_m(0) = \mu m + |m|$, and that $\mathcal{I}_m(x) \sim x + O(1)$ as $x \to \infty$. The eigenvalue condition, Eq. (2.4), reduces to the condition that $\mathcal{I}_m(x)$ should intersect the value $\lambda R$. If $|\mu| < 1$, then the family of curves $\mathcal{I}_m(x)$ will rise without limit for increasing $|m|$, and eventually there will be no solution for negative energies. This corresponds to the intuitive notion that increasing the angular momentum will increase the energy. But notice that if $|\mu| > 1$, then $\mu m + |m|$ can decrease without bound for increasing $|m|$, for one sign of $m$. The set of $\mathcal{I}_m$ which have this behaviour will give rise to an infinite chiral tower of states whose energies are unbounded from below. See Fig. 2 for the graphical argument.

Thus we have proven that the above Hamiltonian is bounded from below only if $|\mu| \leq 1$. A more detailed analysis is required to find the behaviour at the critical point $|\mu| = 1$. It turns out in this case that the Hamiltonian is bounded from below if $\lambda < 0$, has an infinite degeneracy at $E = 0$ if $\lambda = 0$, or is unbounded from below if $\lambda > 0$. For $|\mu| > 1$ the spectrum is necessarily unbounded from below.

Since the states which occur in the infinite chiral tower are described by exponentially growing modified Bessel functions, they actually stick to the edge, with larger angular momentum states having ever finer support near the edge. Such a chiral tower of edge states reminds one of discussions regarding the quantum Hall system [3], and other related systems [11].

## 3. Boundary Conditions and Hall Conductivity From a Spectral Flow

The occurrence of edge states in the effective long-distance field theory for the quantum Hall effect is well understood. Since this effective theory is a Chern-Simons gauge theory, chiral edge states arise from the consideration of gauge transformations which are nonvanishing at the boundary [1]. In the general scheme for edge states in gauge theories [2], this is reasonable. However, another approach is required to understand how edge states can arise from the microscopic electronic theory. Halperin has argued that electronic states localized near the edge of the quantum Hall sample can support a chiral edge current [3]. When the Fermi sea of electronic states is filled, these edge-localized states will be occupied, and the many body ground state will carry a persistent edge current. Our goal in this section is to show how edge states can arise in a single particle picture.

We will now consider the Landau Hamiltonian on the disk,

$$H = -\frac{1}{2M}\left(\vec{p} - e\vec{A}\right)^2, \qquad (3.1)$$

The vector potential in Eq. (3.1) is that of a constant magnetic field normal to the disk. The problem is supplemented by the boundary condition

$$D_r \psi = i\frac{\mu}{R} D_\theta \psi. \qquad (3.2)$$

This is a gauge-covariant form of the boundary condition of Eq. (2.2). In the symmetric gauge, where $A_r = 0$, this condition becomes precisely Eq. (2.2), with $\lambda(\theta)$ being $A_\theta$ restricted to the edge. The Hamiltonian Eq. (3.1), with the boundary condition Eq. (3.2), defines a self-adjoint operator.

It can be shown by direct calculation that the spectrum of this Hamiltonian displays a critical behaviour at $|\mu| = 1$, in exactly the same manner as does the spectrum of the Laplacian on the disk. The details are given in Appendix B. Therefore, for $|\mu| > 1$ a chiral tower of edge states will form, with energies unbounded from below.

Now, the single particle quantum mechanics displayed here and above is ill-behaved for $|\mu| > 1$, since it possesses no ground state. This disease can be cured in time-honored fashion by quantizing the single-particle states as fermionic modes and filling the Fermi sea of states. For the particles on the disk, the sea of states corresponds to the set of one-partice states which have arbitrarily large angular momenta of one sign. This is a chiral sea, and the resulting multi-particle ground state spontaneously breaks parity.

There is a standard argument for displaying the quantized Hall current in this system. According to this argument, one pokes a hole at the origin of the disk and threads a magnetic flux through the hole. An adiabatic change of this flux gives rise to an electric field tangential to the boundary, which satisfies $\nabla \times \vec{E} = \partial B/\partial t$. Changing the central flux by one flux quantum corresponds to a gauge transformation, which should not effect the total system. However, the adiabatic change does have an effect on the bulk and edge states separately; charges flow from the bulk to the edge, and this charge transport is interpreted as the Hall current[4][3]. Explicitly, since the edge excitation is a chiral fermion, there is an anomaly in the edge theory, and the adiabatic addition of one quantum unit of flux results in the creation of an edge excitation. The edge and bulk systems are not separately gauge invariant because of the coupling through the chiral anomaly.

In our second-quantized theory, this behaviour under adiabatic change of the central flux is manifest as a spectral flow across the states of the chiral sea. To see this, note that the angular momentum $m$ always appears in the gauge covariant combination $m + erA_\theta(r)$, which is the gauge-covariant derivative $D_\theta$. The contribution of the central flux to the gauge potential at the boundary is $\Phi/2\pi R$, where $\Phi$ is the central flux. Thus, the change of central flux by one flux quantum, $\Phi_0 = 2\pi/e$, produces a change in the vector potential at the edge of $\delta A_\theta(R) = 1/eR$, and thus an integral change in $m$. The states of the chiral sea therefore shift by one unit.

In usual quantum mechanical situations, spectral flow is rare because of level crossing. In order for a spectral flow to exist, the problem must be effectively one-dimensional so that the spectrum can be ordered and labelled by one parameter. Only then is there a hope that the spectrum of the theory will transform into itself when the control parameter is changed, such as when the central flux in the quantum Hall system is changed by one flux quantum. In our case the edge sector of the theory becomes effectively one dimensional because only one state in each angular momentum sector can bind to the edge. This follows because the delta-function interaction at the edge, which is embodied by our boundary condition, binds precisely one state from each angular momentum sector. The problem thus enjoys an effective dimensional reduction.

Since the quantum mechanical problem is exactly solvable, one can simply plot the wave functions and watch the states rise from and sink into the sea, under the action of the gauge transformation. As a state approaches the top of the sea, its support begins to extend further into the bulk; the wave function expands and the state becomes indistinguishable from a "normal" bulk state.

## 4. How the Boundary Condition Arises From Renormalization

The boundary condition we have imposed at the edge can be interpreted as a local interaction; here we will see exactly how this comes about. Since the boundary condition is chiral, the required interaction at the edge must depend on a choice of direction for the tangent to the edge. We will realize this choice physically, by assigning a current to the edge. The most transparent way to derive the boundary condition from such an interaction is to regulate the current distribution, treating it as a wire of non-zero width. Then a scaling limit can be taken which gives the limiting Hamiltonian that we want.

The interpretation of the boundary condition is simple. A current-carrying wire at the boundary will attract charges from the interior, if they are moving such that their current is aligned with the edge current; anti-aligned currents will be repelled. Furthermore, the aligned states which stick to the edge contribute to the current there, and in this sense the boundary condition can describe a self-consistent solution for the current-carrying wire. We consider the charges of the edge current to be shielded by a neutralizing background so that Coulomb repulsion does not occur.

So we consider a free, charged particle moving in the field of a current which flows at the boundary of the system. Since physics of the boundary is purely local, we can work in the simplest geometry; take the strip $[0,\infty) \times [0, 2\pi L_y)$ and identify the points $y = 0$ and $y = 2\pi L_y$. Suppose a constant current $I$ flows in the $y$ direction, vanishing outside the region $x < a$. This gives rise to a vector potential which, in Landau gauge, can be written as $\vec{A} = A(x)\vec{e}_y$, with

$$A(x) = A(0) + I f(x/a),$$
$$f(x/a) = \begin{cases} x^2/a^2, & x \le a; \\ 1 + \ln(x^2/a^2), & x > a \end{cases} \quad (4.1)$$

The functional form chosen inside the region $x < a$ corresponds to a constant current density. Reducing the problem to one dimension by using the translation invariance in the $y$ direction gives the states $\Psi = \exp(imy/L_y)\psi(x)$, with the family of one-dimensional Hamiltonians

$$H_m = -\frac{1}{2M}\partial_x^2 + V_m(x),$$
$$V_m(x) = \frac{1}{2M}\left[m/L_y + eA(0)\right]^2 \left\{1 + \frac{eI}{m/L_y + eA(0)} f(x/a)\right\}^2. \quad (4.2)$$

Restricting attention to one $m$ sector, we consider the effective potential $V_m(x)$. In the naive limit $a \to 0$, this effective potential goes weakly to zero, meaning that it converges to zero as a distribution acting on smooth functions.

In order to obtain an interesting limit, we must scale the bare coefficients of the potential. Make the choice

$$m/L_y + eA(0) = \frac{L^*}{a}\left(m'/L_y + eA'(0)\right), \qquad (4.3)$$

and also subtract the constant $\delta V_m = \frac{1}{2M}(m/L_y + eA(0))^2$. This constant subtraction has the effect of removing the bulk energy, which is appropriate for consideration of the edge states. The length scale $L^*$ is arbitrary. Then the new effective potential can be written

$$V'_m(x) = \frac{eI}{M}\frac{L^*}{a}\left(m'/L_y + eA'(0)\right)f(x/a) + \frac{e^2 I^2}{2M}f^2(x/a). \qquad (4.4)$$

The second term, with no dependence on the length scale $L^*$, goes weakly to zero in the limit $a \to 0$, but the first term becomes proportional to a delta function at the origin, with a suitable definition of the weak limiting procedure. The details of this limiting procedure are given in Appendix A.

Thus we find that, in a given $m$ sector, the effective potential in the scaling limit is

$$V'_{m'}(x) = L^*\mu\left(m'/L_y + eA'(0)\right)\delta(x), \qquad (4.5)$$

where the constant $\mu$ is given by the appropriate combination of constants introduced in the derivation. This includes an overall arbitrary constant which arises from the renormalization subtraction as discussed in Appendix A.

Fixing the physical effects of this delta-function potential (for example, its effect on the spectrum) is equivalent to fixing a value of $L^*\mu$. Thus we think of $\mu$ as a length-scale dependent coupling, $\mu = \mu(L^*)$, the scaling equation for $\mu$ being

$$\mu(L^*) = \mu(L_0)\frac{L_0}{L^*}. \qquad (4.6)$$

This renormalization group equation for $\mu$ as extremely simple, as is often the case in simple quantum mechanical systems [12]. Our previous analysis showed that the point $\mu(L^*)L^* = 1$ is a *critical* point separating two possible limits. The first limit is the simple scaling limit which gives a theory with no current at the edge, and therefore no spontaneous breaking of parity and no edge states. The second limit is the non-trivial limit that we have exhibited, in which both of the above phenomena occur. In the example of the particle on the disk given above, the natural length scale (renormalization point) is the radius of the disk, $L^* = R$.

Now, this delta-function potential at the edge is precisely what we mean by the boundary condition $\partial_x \Psi = i\mu D_y \Psi$, where the covariant derivative $D_y$ is introduced to give the constant $A'(0)$ the meaning of a vector potential restricted to the edge; compare to standard quantum mechanical calculations for a delta-function potential, implemented by the boundary condition $\partial_x \psi = \kappa \psi$ [13]. Of course, we could as well absorb this constant into the function $\lambda(y)$ which occurs in our generic boundary condition, Eq. (2.2). We could also imagine making $A'$ a function of $y$, although this would require a real-space renormalization, which we have chosen not to implement.

Although the delta-function interaction is strictly local, the large-distance properties of the system enter in an interesting way in the determination of the spectrum. It is well known that the one-dimensional delta-function potential binds precisely one state, no matter how weak the coupling. However, if the delta function is placed inside a system of finite rather than infinite size, the critical coupling for binding moves away from the origin, scaling with the size of the system. As an example where this is easily demonstrated, one can take a particle confined to a box, $[-a, a]$, and insert the delta-function potential at the origin, $\kappa\delta(x)$. Then for $\kappa > 1/a$, a state with energy less than the lowest lying box state will develop. This critical coupling phenomenon for a delta-function potential in a box is precisely the same as that described in the previous Section.

There is one important point which must be mentioned. We have scaled the Hamiltonians for different $m$ separately, without regard to the connection between them. This scaling procedure will generically give rise to a family of Hamiltonians which *do not* represent the dimensional reduction of a higher-dimensional problem. Therefore, as an ingredient in the scaling procedure, we must impose the condition that the final family of Hamiltonians does reassemble itself to become the dimensional reduction of some $H'$. However, there is no obstruction to the realization of this condition since we are free to choose the renormalization prescription separately in each sector. Thus the condition that the family of Hamiltonians reassemble in the indicated manner is a renormalization condition which we can and will impose.

## 5. Contact with a Current Reservoir

It is possible to interpret the boundary condition of Eq.(2.2) in words slightly different from those presented above. In the above Section we saw how this boundary condition can arise from an interaction with a fixed current at the edge. The interaction is such that the charged field contributes to the fixed current and can furthermore become unstable to the attraction toward the edge. We can give the edge current reservoir a more concrete representation by considering the energy of the system. For simplicity, again consider a free particle, with $H = -\Delta$, on the disk. One could equally well consider a more general Hamiltonian involving gauge fields and an arbitrary potential; the argument here will not change.

Integrating by parts, the expectation value of the energy can be written

$$\begin{aligned}\langle E \rangle &= \frac{1}{2M}\int_D |\nabla\psi|^2 - \frac{1}{2M}\int_{\partial D} \bar\psi \partial_r \psi \\ &= \frac{1}{2M}\int_D |\nabla\psi|^2 - \frac{1}{2M} i\frac{\mu}{R}\int_{\partial D} \bar\psi \partial_\theta \psi.\end{aligned} \quad (5.1)$$

We assume the state is sufficiently smooth for this operation. In the second line we have used the boundary condition (2.2). Suppose, without loss of generality, that the state $\psi$ is an eigenstate of angular momentum, $\psi \propto e^{im\theta}$. Then we can write

$$\langle E \rangle = \frac{1}{2M}\left[\int_D |\nabla\psi|^2 + m\mu 2\pi |\psi(R)|^2\right]. \quad (5.2)$$

In this equation, we recognize the second term as having the form $J \cdot \omega$, which is the energy of a body with angular momentum $J$ spinning with angular velocity $\omega$. The angular velocity and angular momentum are

$$\begin{aligned}\omega &= \frac{\mu}{MR^2}, \\ J &= \pi R^2 m |\psi(R)|^2.\end{aligned} \quad (5.3)$$

The interpretation of $J$ as an angular momentum depends on a "rigidity" condition, since the density which appears is actually evaluated at the edge rather than averaged over the bulk. This is appropriate for the superfluid analogy which we are cultivating at this point.

In slightly different words, the angular velocity is a chemical potential for angular momentum, and thus we see that $\mu$ is proportional to the chemical potential for angular momentum. When the particle is charged, the angular momentum carries a current, and $\mu$ has the interpretation of a chemical potential for the system interacting with a current reservoir at the boundary.

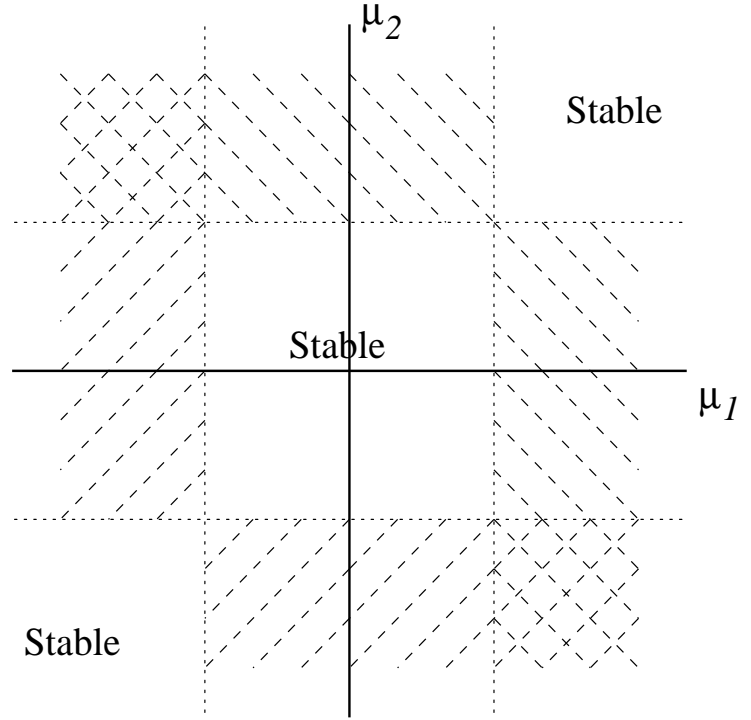

Figure 3. Phase diagram for the case of the Laplacian on the annulus. Regions marked "Stable" correspond to systems with spectra that are bounded from below. Hatched regions correspond to systems where states of one chirality become unstable. The doubly hatched regions correspond to systems where states of both chiralities become unstable. The dotted lines separating the various regions occur at the critical values $|\mu_i| = 1$.

## 6. Remarks About Cancelling Boundaries

As discussed above, a spectral catastrophe occurs for the Laplacian on the disk when $|\mu|$ becomes large enough, with the spectrum becoming unbounded from below. One can imagine filling the negative energy chiral sea of one-particle states, also as discussed above. One can also ask what happens when two (or more) disconnected boundary components are present. It turns out that extra boundaries can stabilize the system. We will discuss this second point now.

Consider again the example of the Laplacian, but on an annulus, with circular boundaries at $R_1$ and $R_2$, $R_1 < R_2$. The boundary conditions are

$$\begin{aligned} -\partial_r \Psi &= i\mu_1 \tfrac{1}{R_1} \partial_\theta \Psi, & r &= R_1, \\ \partial_r \Psi &= i\mu_2 \tfrac{1}{R_2} \partial_\theta \Psi, & r &= R_2. \end{aligned} \quad (6.1)$$

The sign in the first equation is a convention and accounts for the orientation of the boundary. By direct calculation, one arrives at the phase diagram shown in Fig. 3. The interesting point here is the existence of a stabilized spectrum for arbitrarily large values of $\mu_i$, along the direction of $\mu_1 \simeq \mu_2$.

We have seen that if the angular velocity at one edge is greater than a certain quantum unit, then the naive particle vacuum becomes unstable. However, the calculation on the annulus shows that one spinning boundary can cancel the ill effects of another, if the boundaries are properly matched. Therefore, the naive vacuum can be stabilized by the production of an interior *vortex* inside the disk.

Such production of vortices is a well-known phenomena in the theory of superfluid Helium. The critical angular velocity for the production of vortices in a spinning bucket of superfluid is $\omega_c \simeq \hbar/Md^2$, where $M$ is the particle mass and $d$ is a characteristic size [14]. Our critical value of $|\mu| = 1$ corresponds to a critical angular velocity of $\hbar/(MR^2)$ The agreement basically follows from dimensional analysis. In a more sophisticated superfluid calculation one must take account of the interaction of the interior vortex with its image, introducing a logarithmic correction which depends on the size of the system [15].

If the fundamental excitations are quantized as bosons, as in the case of the superfluid, Fermi statistics cannot be used to define a vacuum sea, and the naive vacuum should be stabilized by some other mechanism. There is no choice but to puncture the system at one or more places and insert an appropriate number-current loop at the puncture, representing a spinning boundary. This is in contrast to the fermionic case where it was possible to fill the states of the naive vacuum.

## 7. Discussion

The physics presented here is not essentially new. One sees here the appearance of the standard ideas associated with the quantum Hall effect. What is new here is the realization of these ideas in terms of simple Schrödinger quantum mechanics supplemented with a novel boundary condition. Perhaps this is testament to the universal nature of the ideas surrounding the quantum Hall effect. The merit of our picture is its simplicity. Not only does one have the machinery of the anomaly easily displayed, but one also has the simple picture of Schrödinger quantum mechanics in which this machinery resides.

One can ask if the production of vortices plays a role in the quantum Hall system, as it did in the superfluid case described above. In the superfluid case one should second-quantize the particles as bosons, and therefore an instability of the ground state is not acceptable. In that case, the only recourse for the system is a production of interior vortices. But in the Hall system one imagines that the particles should be quantized as fermions, with the formal second-quantized ground state representing the filled Fermi sea in the manybody theory. Then the question of interior vortex production becomes a more detailed one. Vortices may or may not appear, possibly depending on energetic considerations.

The idea that vortex excitations can arise in the Hall fluid is well known [16][17][18][19]. Such excitations are called magneto-rotons, and they are expected to occur in systems with a fractionally filled Landau level. Some support for this idea comes from agreement with numerical simulations, assuming the Laughlin wave functions correctly describe the ground state [18][19]. The current distribution for a magneto-roton state is circular, with vanishing divergence [19]. The similarity to an edge current distribution at the boundary of a puncture is clear.

Just as in the case of the superfluid, the one-particle quantum mechanics cannot be used to describe the collective vortex excitation in the Hall system. However, again following the superfluid case, the effect of the vortex insertion on the one-particle states can be described by a suitable boundary condition at the vortex puncture.

## 8. Acknowledgements

We are indebted to A. P. Balachandran for suggesting the problem and for a critical reading of the manuscript. We also acknowledge conversations with L. Chandar, A. Momen, and especially D. Sen. This work was supported by the U.S. Department of Energy under contract No. DE-FG02-85ER40231

## Appendix A. Weak Limit of the Effective Potential

In the discussion of Section 2, we used a limiting procedure to produce a non-trivial effective potential in the limit of a thin current sheet. This limiting procedure requires some definition.

Consider a function $h(w)$, which is rapidly decreasing over $[0,\infty)$. Then we have, in the sense of a weak limit,

$$\lim_{a\to 0} \frac{c}{a} h(x/a) = c\,\delta(x). \tag{A.1}$$

To see that this is so, let $g(x)$ be an arbitrary smooth function of rapid decrease, real analytic in a neighbourhood of the origin. Then

$$\lim_{a\to 0}\int_0^\infty dx\, \frac{h(x/a)}{a} g(x) = \lim_{a\to 0}\int_0^\infty dw\, h(w)g(wa)$$
$$= g(0)\int_0^\infty dw\, h(w) + O(a). \tag{A.2}$$

The condition of analyticity can obviously be dropped.

In the case at hand, Eq. (4.4), we have a function $h(w)$ which is not decreasing and is, in particular, not integrable over $[0,\infty)$. In order to define the weak limit, we must modify the function at large distances by making an appropriate subtraction. Consider the limit

$$d(x) = \lim_{L\to\infty}\lim_{a\to 0}\frac{1}{a}\left[h(x/a) - K(L/a)\right]\theta(x<L), \tag{A.3}$$

which defines a distribution $d(x)$; the limit $a\to 0$ is taken holding the quantity $L/a$ fixed. Let $g(x)$ be a test function as before. By algebra similar to the above, we have

$$d[g] = \lim_{\eta\to\infty} g(0)\left[\int_0^\eta dw\, h(w) - K(\eta)\eta\right]. \tag{A.4}$$

By choosing $K(\eta) = \eta^{-1}\int_0^\eta h + C\eta^{-1}$, it is easily seen that $d(x) = C\delta(x)$. The renormalization of the long-distance behaviour has introduced an arbitrary constant, $C$.

Of course, a choice is made here in the definition of the regulator. This definition corresponds to a subtraction of the "average" value of the potential at infinity, leaving only the structure near the origin, which collapses to a delta function. The same algebra shows that the second term in Eq. (4.4) does in fact go weakly to zero, as claimed in Section 3.

## Appendix B. Spectrum of the Landau Hamiltonian

In this Appendix we provide the detailed calculations which display the spectral instability for the Landau Hamiltonian with the boundary condition (3.2), on a disk of radius $R$. Let $\xi = |eB|r^2/2$ and $\bar\xi = \xi(R)$. The radial dependence of the wave functions is governed by the confluent hypergeometric equation, and the basic solutions to the problem have the form

$$\psi = e^{im\theta}\exp(-\xi/2)\xi^{|m|/2}\,{}_1F_1(\alpha, 1+|m|;\xi). \tag{B.1}$$

The parameter $\alpha$ is determined by application of the boundary conditions, and it is related to the energy by

$$\alpha = \frac{1+|m|}{2} - \frac{|m|}{2}\,\mathrm{sgn}(eB) - \epsilon \tag{B.2}$$

where $\epsilon = 2ME/|eB|$. The boundary condition in the symmetric gauge is

$$\partial_r\psi = \left[\frac{eBR\mu}{2} + i\mu\frac{1}{R}\partial_\theta\right]\psi \quad\text{at } r = R. \tag{B.3}$$

Substituting the wave function into this condition gives the transcendental equation $\mathcal{I}_m(\mu,\epsilon) = 0$, where

$$\mathcal{I}_m(\mu,\epsilon) \equiv |m| + \mu m - \bar\xi(1 + \mu\,\mathrm{sgn}(eB)) + \frac{2\alpha\bar\xi}{1+|m|}\frac{{}_1F_1(1+\alpha, 2+|m|;\bar\xi)}{{}_1F_1(\alpha, 1+|m|;\bar\xi)}. \tag{B.4}$$

We wish to solve the equation for $\epsilon$.

Suppose that $eB > 0$. Then we have

$$\mathcal{I}_m(\mu, \epsilon) = \begin{cases} |m|(1+\mu) - \bar{\xi}(1+\mu) + \dfrac{\bar{\xi}(1-2\epsilon)}{1+|m|} \dfrac{{}_1F_1(3/2-\epsilon, 2+|m|; \bar{\xi})}{{}_1F_1(1/2-\epsilon, 1+|m|; \bar{\xi})}, & m \geq 0; \\ |m|(1-\mu) - \bar{\xi}(1+\mu) + \dfrac{\bar{\xi}(1+2|m|-2\epsilon)}{1+|m|} \dfrac{{}_1F_1(1/2+\epsilon, 2+|m|; -\bar{\xi})}{{}_1F_1(1/2+\epsilon, 1+|m|; -\bar{\xi})}, & m < 0. \end{cases}$$
(B.5)

A similar expression exists for the case $eB < 0$. Clearly the $\mathcal{I}_m$ are analytic functions of $\epsilon$ wherever the denominator is non-vanishing. In particular they are analytic for $\epsilon < 1/2$. Also, the $\mathcal{I}_m$ cannot decrease without bound for $\epsilon < 1/2$, since

$$\mathcal{I}_m(\mu, \epsilon) \geq |m| + \mu m - \bar{\xi}(1 + \mu \operatorname{sgn}(eB)) \qquad \text{for } \epsilon \leq 1/2.$$
(B.6)

Furthermore, we have the asymptotic behaviour for large negative $\epsilon$,

$$\mathcal{I}_m(\mu, \epsilon) \sim 2\sqrt{|\epsilon|\bar{\xi}} + \mu m - \frac{1}{2} - \bar{\xi}(1+\mu\operatorname{sgn}(eB)) + O(|\epsilon|^{-1/2}) \qquad \epsilon \to -\infty.$$
(B.7)

Since a non-constant analytic function can assume any given value at most finitely many times in a compact set, we immediately have that there at most finitely many negative energy modes when $\mu < 1$. As we have noted above, no poles of the $\mathcal{I}_m$ lie to the left of $\epsilon = 1/2$. Given the fact that $\partial \mathcal{I}_m / \partial \epsilon < 0$, this implies that each value of $m$ can give rise to at most one negative energy state. This holds regardless of the values of $\mu$, $eB$, or $\bar{\xi}$.

If $|\mu| > 1$, then the plots of $\mathcal{I}_m$ will descend down the $(\epsilon, \mathcal{I})$ plane, for those $m$ with sign fixed by the sign of $\mu$. States with the other sign of $m$ will give rise to at most finitely many negative energy states, by the same argument as given above for the case $|\mu| < 1$. We see that the mechanics of the argument are precisely analogous to those of the argument given in Section 2 for the Laplacian on the disk.